\documentclass[]{article}

\usepackage[dvips]{graphicx}
\usepackage{amssymb}

\title{Orienting Ion-Containing Block Copolymers
Using Ac Electric Fields}
\author{Yoav Tsori\footnote{e-mail: yoav.tsori@espci.fr},
Fran\c{c}ois Tournilhac ~and Ludwik Leibler\\
Laboratoire Mati\`ere Molle \& Chimie (UMR 167)\\
ESPCI, 10 rue Vauquelin,
 75231 Paris CEDEX 05, France}

\date{23/6/2003}

\newcommand{\eps} {\varepsilon}

\begin{document}

\maketitle

\begin{abstract}
We consider orientation mechanisms for block copolymers in an
electric field. Theoretical and experimental studies have shown
that nonuniformity of the dielectric constant gives rise to a
preferred orientation of the melt with respect to the applied
field. We show that the presence of ions, as found in anionically
prepared copolymers, may increase the alignment effect markedly.
Time-varying (ac) and static (dc) fields are considered within a
unified framework. We find that orientation of block copolymers
can in principle be achieved without a dielectric contrast if
there is a mobility contrast. The presence of ions is especially
important at small field frequencies, as is in most experiments.
Unlike the no-ions case, it is found that orienting forces depend
on the polymer chain lengths. The mobile-ions mechanism suggested
here can be used to reduce the magnitude of orienting fields as
well as to discriminate between block copolymers of different
lengths.

\end{abstract}

\section{Introduction}

Numerous applications of ordered mesophases in soft materials have
emerged in recent years. Few examples are antireflection
coatings,$^{\cite{steiner1}}$ photonic band-gap materials
$^{\cite{colloidbgp}}$ and waveguides,$^{\cite{fink1}}$ and
ordered arrays of thin metallic wires.$^{\cite{russell1}}$ In
these systems it is crucial that the crystalline domains are
oriented in a specific direction. Several techniques have been
used to this end. Mechanical shear is effective in aligning bulk
systems and in annealing defects.$^{\cite{fredrickson1}}$. Static
(dc) electric fields are widely used in thin films of block
copolymers (BCP) to control the film
morphology.$^{\cite{russell2,krausch1,onuki95,PW99,TA02}}$ In this
paper, we focus on the alignment mechanism of a BCP melt in
time-varying (ac) electric fields.

The pioneering work of Amundson and Helfand and co-workers
$^{\cite{AH93,AH94}}$ showed that diblock copolymers (composed of
two different chemical sequences tethered at one point) in the
lamellar phase can be aligned parallel to an external dc electric
field. The driving force in this alignment is the ``dielectric
mechanism''. In this mechanism, there is an electrostatic free
energy penalty for having dielectric interfaces perpendicular to
the external field. Thus, lamellae (or cylinders) will tend to
align parallel to the field.

However, most studies to date have given little or no attention to
the ions in BCP. These ions naturally exist in many polymeric
systems and, in particular, in anionically prepared BCP. To see
this consider a typical polymerization reaction initiated with one
butylithium (BuLi) group. At the end of polymerization and after
rinsing with water, the Li and OH ions form a lithium hydroxide
(LiOH) pair. Hence, unless special measures are taken (e.g., final
purification by continuous methanol extraction) ions exist in
abundance in the melt, i.e., one ion pair per polymer chain. Some
fraction of these ions are dissociated (either thermally or
because of a high polymeric solvation ability), and those positive
and negative charges move under the influence of the field. This
motion, in turn, changes the field and exerts forces on the melt.

This paper generalizes the results of previous studies to a BCP
melt containing dissociated mobile ions in time-varying fields. As
we will see below, in these systems there are strong orienting
forces. Forces occur by the presence of a {\it conductivity} or
{\it mobility} contrast $^{\cite{Nitzan}}$ and can occur even
without dielectric contrast. The details of our calculation show
that in many common BCP melts and under normal circumstances the
relative strength of the orienting forces due to ions leads to
substantial reduction in the magnitude of external
fields.$^{\cite{TTAL}}$

\section{Electric field in the melt}

To find what are the orienting forces on the sample, we need to
calculate the distribution of electric field inside a nonuniform
polymer system such as block copolymers in one of the ordered
phases.$^{\cite{leibler1,schick1}}$ It is therefore implicitly
assumed that the electric field relaxes faster than any other
process, thereby instantaneously adjusting to the BCP morphology.
We adopt a description neglecting the microscopic polymer
chemistry, architecture, etc., and  focus on coarse-grained
variables. These are the dielectric constant $\eps({\bf r})$, the
ionic mobility $\mu({\bf r})$, number density $\rho^\pm({\bf r})$
for the two ionic species (positively and negatively charged,
respectively), and electric field ${\bf E}({\bf r})$. The average
ionic densities $\rho_0^\pm=\langle \rho^\pm({\bf r})\rangle$ can
be controlled by sample doping, etc. The first two quantities are
material properties and are given solely as a function of the BCP
density. It should be noted that the BCP melt considered by us is
different from the ``leaky dielectric'' model of Taylor and
Melcher $^{\cite{taylor1}}$ for two conducting dielectric fluids.
Here the chain behavior is elastic and there is no hydrodynamic
flow. In the weak-segregation regime employed below (high
temperature region of the ordered phases), the chains are weakly
stretched and it is appropriate to neglect deviations from
anisotropy of the dielectric and mobility tensors and take $\eps$
and $\mu$ to be scalars. For example for an A/B diblock copolymer
with an average fraction $f$ of A monomers, the order parameter
$\phi$ is defined as the deviation of the A-monomer concentration
from its average value, $\phi({\bf r})=\phi_A({\bf r})-f$. For
symmetric ($f=1/2$) diblock copolymers
\begin{eqnarray}
\eps({\bf r})=\phi({\bf r})(\eps_A-\eps_B)+\eps_0
\end{eqnarray}
where $\eps_A$ and $\eps_B$ are the dielectric constants of the A
and B monomers, respectively, and $\eps_0=\frac12(\eps_A+\eps_B)$
is the material average dielectric constant. The quantities to be
found are $\rho^\pm({\bf r})$ and ${\bf E}({\bf r})$.
\begin{figure}[h!]
\begin{center}
\includegraphics[scale=0.55,bb=60 500 540 720,clip]{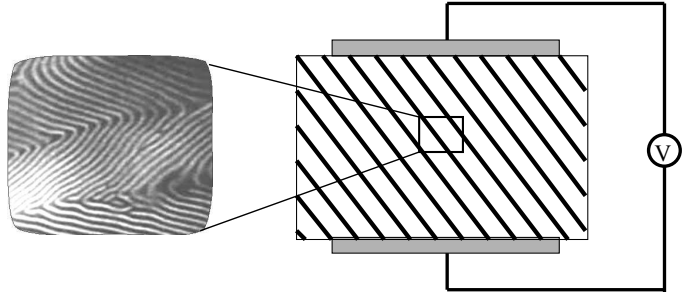}
\end{center}
\caption{Schematic illustration of a block copolymer melt
(diagonal lines) between two flat and parallel electrodes (dark
gray). Inset shows a typical morphology as revealed by
transmission electron microscopy.}
\end{figure}

The BCP melt is put between two flat parallel plates connected to
an oscillatory voltage supply of frequency $w$, $V(t)=V_0e^{iwt}$,
as is schematically seen in Figure~1. The $x$ axis is
perpendicular to the plates. We concentrate on the so-called
``weak-segregation'' limit where the polymer density deviations
from the average value are small and a linearization procedure
suffices. Then our goal is to deduce what is the best orientation
of the dielectric-conducting material with respect to the external
field ${\bf E}_0$. Such a conducting system is dissipative, and
hence, we look for mechanical stability instead of thermodynamic
equilibrium.

There are four length scales in the system. The first is $d_0$,
the natural crystal periodicity, $d_0\simeq 50$-$200$ nm. For
weakly a segregated diblock copolymer of $N$ monomers , $d_0\sim
N^{1/2}$. The other length scale is $l_{\rm drift}\equiv \pi e\mu
E/w$, the drift distance an ion of charge $e$ and mobility $\mu$
undergoes in a field $E$ in a time equal to half of the
oscillation period. We call the high-frequency regime the regime
where $l_{\rm drift}\ll d_0$ ; the low-frequency regime is when
$l_{\rm drift}\gg d_0$. The third length is the wavelength of
light, and is given by $l_{\rm light}=2\pi c/(w\sqrt{\eps})$. In
most circumstances $l_{\rm light}$ is larger than the two other
lengths, $l_{\rm light}\gg d_0$, $l_{\rm light}\gg l_{\rm drift}$.
The last length scale is the system size $L$.
\begin{center} \footnotesize
\begin{tabular}{|c|c|c|}
\hline
   & PMMA & PS \\ \hline
  sample aspect & transparent & hazy \\ \hline
  birefringence & no & strongly \\ \hline
  number of LiOH & $6.6\cdot 10^{14}$ & $2.6\cdot 10^{14}$\\
  ions per m$^3$ & &  \\ \hline
  fraction of dissociated & $3\cdot 10^{-5}$ & below\\
  ion plirs &  & experimental error \\ \hline
  mobility & $4.6\cdot 10^5$ & below\\
  ${\rm [m}^2{\rm /Joule~sec]}$ &  & experimental error \\ \hline
\end{tabular}
~\\
\end{center}
\footnotesize{Table 1. Data Collected from Dc Conductivity
Measurements of PMMA and PS Doped with LiOH Ions at
$160^\circ$.}\normalsize
~\\

There are three energy scales in the system. One is the
electrostatic energy stored in unit volume of the dielectric
material, $U_{\rm dielec}\equiv \eps E^2$. The second is the Joule
heating per unit volume in one field cycle, $U_{\rm Joule}\equiv
2\pi E^2\sigma/w$. The third energy is the thermal energy $kT$. As
the frequency of field decreases Joule heating becomes more
dominant. The Table shows some of the data collected by us on
conductivity of poly(methylmethacrylate) (PMMA) and polystyrene
(PS) doped with Li ions. These polymers constitute one of the best
studied pairs in diblock copolymers. Let us estimates $U_{\rm
dielec}$ and $U_{\rm Joule}$ for pure PMMA polymer. In anionically
prepared polymers there is approximately one LiOH group per chain.
To evaluate how many of these ions participate in the alignment
process we conducted dc conductivity measurements on pure PMMA
doped with a known amount of LiOH. The samples were sandwiched
between two conducting flat indium Tin Oxide coated glass
substrates at $160$ $^\circ$ C. Integration of the current allowed
us to deduce that a fraction of $3\times 10^{-5}$ of these ions
are dissociated and {\it mobile} in PMMA. We concluded that the
mobile ion number density is $\rho_0\simeq 3\times 10^{22}$
$m^{-3}$. Li ion mobility in PMMA is $\mu_0\simeq 4.6\times 10^5$
$m^2/$(J sec), giving an average conductivity
$\sigma_0=e^2\rho_0\mu_0\simeq 3.5\times 10^{-10}$ C/(m V s). For
frequencies of $50$ Hertz we therefore have $\sigma_0/w\simeq
1.1\times 10^{-12}$ C/(m V). On the other hand, $\eps_{_{\rm
PMMA}}\simeq 6\times 8.85~10^{-12}$ C/(m V), so we find that
$2\pi\sigma_0/w\simeq 0.13\times \eps_{_{\rm PMMA}}$ (or $U_{\rm
Joule}/U_{\rm dielec}\simeq 0.13$). This estimate shows us that we
should expect free ions (conductive behavior) to start playing an
important role at {\it low} frequencies.

An important question is whether the heating caused by moving ions
can cause problems in the experimental setup. The total heat
created per unit time in the volume is $\sim L^3\sigma E^2$ ($L$
is the system size). This heat should be compared with the
integrated heat flux out of the melt, $\sim L^2\kappa\nabla T$,
where $\kappa$ is the heat conductivity, and the temperature
gradient can be approximated by $\nabla T\simeq \Delta T/L$. In
most setups the sample temperature variation is controlled to
within $\Delta T\simeq 1~{\rm K}$. Concentrating on PMMA where
$\kappa\simeq 0.19$ $W/(m {\rm K})$ $^{\cite{polyHB}}$ and
considering a sample of size $L=1$ cm and applied field $E=1000$
$V/{\rm cm}$ we find that the ratio between the heat created to
the heat that can flow outside of the sample is $1.8\times
10^{-3}$ and thus heating does not present a problem. In the case
of a smaller sample of $L=1$ $\mu$m this ratio becomes even
smaller, $1.8\times 10^{-7}$ (small samples have a large surface
area ).

Below we employ a formal derivation using Maxwell's equations to
find the electric field distribution in nonhomogeneous media.
While we will be interested in relatively low frequencies relevant
to experiments, the results we obtain are general. Maxwell's
equations in the CGS system of units are:
\begin{eqnarray}
{\bf \nabla}\times{\bf E}+\frac1c\dot{\bf B}&=&0\label{m1}\\
{\bf \nabla}\times{\bf B}-\frac{\eps}{c}\dot{\bf E}&=&
\frac{4\pi}{c}{\bf J}~,~~~~~~~~~~~~~~~ {\bf
J}=e^2\left(\mu^+\rho^++\mu^-\rho^- \right){\bf E} \label{m2}
\end{eqnarray}
${\bf B}$ is the magnetic field, ${\bf J}$ is the current density,
and $\mu^+$ and $\mu^-$ are the mobilities of the positive and
negative ions, respectively.  The {\it number} densities of
positive and negative ions are denoted by $\rho^+$ and $\rho^-$,
respectively, all assumed monovalent. We put eq \ref{m1} into eq
\ref{m2} and add two continuity equations for the positive and
negative ions to have three governing equations:
\begin{eqnarray}
-{\bf \nabla}({\bf \nabla}{\bf E})+\nabla^2{\bf
E}-\frac{\eps}{c^2}\ddot{\bf E}&=&\frac{4\pi e^2}{c^2}\left[\mu_
+\left(\dot{\rho}^+{\bf E}+\rho^+\dot{\bf
E}\right)+\mu^-\left(\dot{\rho}^-{\bf E}+\rho^-\dot{\bf
E}\right)\right]\label{gov_eqn1}
\\
\frac{\partial \rho^+}{\partial t}+{\bf
\nabla}\left(e\mu^+\rho^+{\bf E}\right)
&=&0\label{gov_eqn2}\\
\frac{\partial \rho^-}{\partial t}- {\bf \nabla}\left(e\mu^-\rho^-
{\bf E}\right) &=&0\label{gov_eqn3}
\end{eqnarray}

We are interested in the weak-segregation regime of the phase
diagram, which is the high-temperature region of the various
ordered phases. The weak segregation is more relevant to
experiments because the melt is not so viscous as it is for strong
segregations (low temperatures). The copolymer density is slowly
varying and the interfaces between domains are smooth.  As we show
below, this allows to replace the complicated task of numerically
solving eqs (\ref{gov_eqn1}), (\ref{gov_eqn2}), and
(\ref{gov_eqn3}) by an approximate analytical procedure. In this
scheme, the material quantities $\eps({\bf r})$ and $\mu({\bf r})$
are given by the expansions $\eps=\eps_0+\eps_1$ and
$\mu^+=\mu^-=\mu_0+\mu_1$, with $\eps_1\ll \eps_0$ and $\mu_1\ll
\mu_0$, where for simplicity we assume hereafter that positive and
negative ions have the same mobility. The charge density and
electric field can be similarly written as
$\rho^\pm=\rho^\pm_0+\rho^\pm_1$ and ${\bf E}={\bf E}_0+{\bf
E}_1$.

The zero-order approximation of eqs \ref{gov_eqn1},
\ref{gov_eqn2}, and \ref{gov_eqn3} gives the field ${\bf E}_0$:
${\bf E}_0=\tilde{{\bf E}}_0e^{i{\bf k}_{\rm light}(w){\bf
r}+iwt}~+c.c.$, where $c.c.$ stands for the complex conjugate
operation and the $k$-vector obeys the dispersion relation:
\begin{eqnarray}
k_{\rm light}^2(w)&=&\frac{\eps_0w^2-4\pi
ie^2w\mu_0(\rho_0^++\rho_0^-)}{c^2}\nonumber\\
&=&\frac{w^2}{c^2}\left(\eps_0-8\pi i\sigma_0/w\right)\ll k^2
\end{eqnarray}
$k_{\rm light}$ is parallel to the electrodes (and perpendicular
to ${\bf E}_0$). The second line above is correct for neutral
media with $\rho_0^\pm=\rho_0$ and where the average conductivity
$\sigma_0$ has been identified as $\sigma_0=e^2\mu_0\rho_0$. In
the inequality $k_{\rm light}^2\ll k^2$, $k$ is inversely
proportional to the characteristic BCP length scale. The
inequality is valid because even at visible frequencies (e.g., red
light, $f=4.3\times 10^{14}$ Hz) the wavelength $\lambda=700$ nm
is larger than any structure size in block copolymers ($l_{\rm
light}\gg d_0$); we are interested in much smaller frequencies.

The electric field in a conducting media decays exponentially but
it can be regarded as approximately uniform (with less than $10$\%
decrease across the sample) if ${\rm Im}(k_{\rm light}) L<0.1$,
where $L$ is the system size. Below we concentrate on two regimes:
the first is where Joule heating $U_{\rm Joule}$ in one field
cycle is larger than the energy $U_{\rm dielec}$ stored in the
dielectric, $2\pi E^2\sigma/w\gg \eps E^2$. In this regime ${\bf
E}_0$ is approximately uniform if $0.1 c/L> \sqrt{4\pi\sigma_0w}$.
The second regime is where $U_{\rm Joule}\ll U_{\rm dielec}$, and
it is then required that $0.1 c/L> 4\pi \sigma_0/\sqrt{\eps_0}$.
When these requirements are met the field is approximately uniform
and $\tilde{{\bf E}}_0={\bf E}_0$.

We continue by writing the deviations from uniform
quantities $\eps_1$, $\mu_1$, ${\bf E}_1$ and $\rho^\pm_1$
by using the Fourier representation
\begin{eqnarray}
\eps_1({\bf r})&=&\sum_{\bf k}\eps_{\bf k}e^{i{\bf k}\cdot{\bf
r}}~+c.c.~,~~~~~~~~~~~~~~~\mu_1({\bf r},t)=\sum_{\bf k}\mu_{\bf
k}e^{i{\bf k}\cdot{\bf
r}}~+c.c.\nonumber\\
{\bf E}_1({\bf r},t)&=&\sum_{\bf k}{\bf E}_{\bf k}e^{i{\bf
k}\cdot{\bf r}+iwt}~+c.c.~,~~~~~~~~~
 \rho_1({\bf r},t)=\sum_{\bf k}\rho_{\bf
k}e^{i{\bf k}\cdot{\bf r}+iwt}~+c.c.\label{fourier}
\end{eqnarray}
We are interested in a mesoscopic system where many modes
participate in the sum; For an infinite system the sums over all
existing $k$-modes become integrals. Since the electric potential
has to comply with the boundary conditions on the electrodes,
${\bf E_k}$ is nonzero only for a restricted set of $k$'s obeying
$kL=2n\pi$, with integer $n$.

The Fourier representation of eq \ref{fourier} is substituted in
eqs \ref{gov_eqn1}, \ref{gov_eqn2} and \ref{gov_eqn3}. After some
algebra one finally has
\begin{eqnarray}\label{gen_E}
{\bf E_k}&=&\frac{C-B\mu_k{\bf k}\cdot{\bf E}_0}{k_{\rm
light}^2-k^2}\left[{\bf E}_0-\frac{{\bf k}\cdot{\bf E}_0}{k_{\rm
light}^2+B\mu_0{\bf k}\cdot{\bf E}_0}\left({\bf
k}+B\mu_0{\bf E}_0\right)\right]\nonumber\\
&=&\frac{C-B\mu_k{\bf k}\cdot{\bf E}_0}{\left(k_{\rm
light}^2-k^2\right)\left(k_{\rm light}^2+B\mu_0{\bf k}\cdot{\bf
E}_0\right)}\left[k_{\rm light}^2{\bf E}_0-({\bf k}\cdot{\bf
E}_0){\bf k}\right]
\end{eqnarray}
with $B$ and $C$ given by
\begin{eqnarray}
B&=&-\frac{8\pi
e^2iw\mu_0}{c^2}\frac{w(\rho_0^--\rho_0^+)/e+\mu_0{\bf k}\cdot
{\bf E}_0(\rho_0^++\rho_0^-)}{w^2/e^2-(\mu_0{\bf k}
\cdot{\bf E}_0)^2}\nonumber\\
C&=&\frac{4\pi e^2iw}{c^2}\mu_k(\rho_0^++\rho_0^-)-\frac{\eps_k
w^2}{c^2}\label{C-B}
\end{eqnarray}

\section{Orienting with electric field}

At this point we have the general expression for the field in the
block copolymer. When there are no ions in the system,
$\rho^\pm=0$, eq \ref{gen_E} reduces to the expression used by
Amundson and Helfand. We adhere hereafter to overall charge
neutrality, $\rho_0^+=\rho_0^-\equiv \rho_0$, and to the
experimentally relevant regime where $k_{\rm light}\ll k$, or
equivalently, $l_{\rm light}\gg d_0$. The Fourier component of the
field is
\begin{eqnarray}
{\bf E_k}&=&-\frac{T\mu_k-\eps_kw^2}{T\mu_0-
\eps_0w^2}\frac{{\bf k}\cdot{\bf E}_0}{k^2}{\bf k} \label{small_k0}\\
T&=&8\pi e^2iw\rho_0\frac{(w/e)^2+(\mu_0{\bf k}\cdot{\bf
E}_0)^2}{(w/e)^2-(\mu_0{\bf k}\cdot{\bf E}_0)^2}\nonumber
\end{eqnarray}
Compared to the classical expression, here the ratio
$\eps_k/\eps_0$ is replaced by a direction-dependent quantity
(containing ${\bf k}{\bf E}_0$ via $T$). This fact is significant
because it means that the field and the orienting force depend on
the periodicity of the BCP melt, whereas in the purely dielectric
case (no ions) the field depends on the direction of the
wavevector ${\bf k}$ but not on its absolute value. Hence in
contrast to previous predictions, two melts with the same chemical
sequence but different chain lengths are oriented differently. For
low frequencies (see below), longer polymers (smaller $k$'s) are
expected to feel a stronger aligning force that shorter polymers.

The discussion continues with two distinct cases,
corresponding to low and high frequencies.

$\bullet$ Low frequency limit: $w/e\ll \mu_0{\bf k}{\bf E}_0$
($l_{\rm drift}\gg d_0$) The meaning of this limit is that the
drift distance an ion undergoes in the external field is much
larger than any BCP length scale (crystal periodicity). In this
limit the directionality term ${\bf k}{\bf E}_0$ in $T$ is lost
and we have
\begin{eqnarray}\label{lfl_eqn}
{\bf E_k}&=&-\frac{8\pi e^2iw\rho_0\mu_k+\eps_kw^2}{8\pi
e^2iw\rho_0\mu_0+\eps_0w^2}\frac{{\bf k}\cdot{\bf E}_0}{k^2}{\bf
k}
\end{eqnarray}

$\bullet$ High frequency limit: $w/e\gg \mu_0{\bf k}{\bf E}_0$
($l_{\rm drift}\ll d_0$) Here the ionic drift distance in the
electric field is much smaller than the BCP  length scale, so
charges are almost fixed in their positions. In this limit too the
directionality term ${\bf k}{\bf E}_0$ is lost, yielding ${\bf
E_k}$
\begin{eqnarray}
{\bf E_k}&=&-\frac{8\pi e^2iw\rho_0\mu_k-\eps_kw^2}{8\pi
e^2iw\rho_0\mu_0-\eps_0w^2}\frac{{\bf k}\cdot{\bf E}_0}{k^2}{\bf
k}
\end{eqnarray}

\section{Force and torque}

Systems with currents are generally dissipative and therefore the
steady-state of the system is not given by free energy
minimization, which is ill-defined under such circumstances.
Rather, the system state is given by mechanical stability. The
total torque has an effect on the dielectric material (bound
charges) as well as on free charges, ${\bf L}={\bf L}_{\rm
dielec}+{\bf L}_{\rm ions}$. To understand the origin of torque in
a nonhomogeneous system in electric field, consider the most
simple case of a lamellar system depicted in Figure~2.
\begin{figure}[h!]
\begin{center}
\includegraphics[scale=0.55,bb=80 420 450 740,clip]{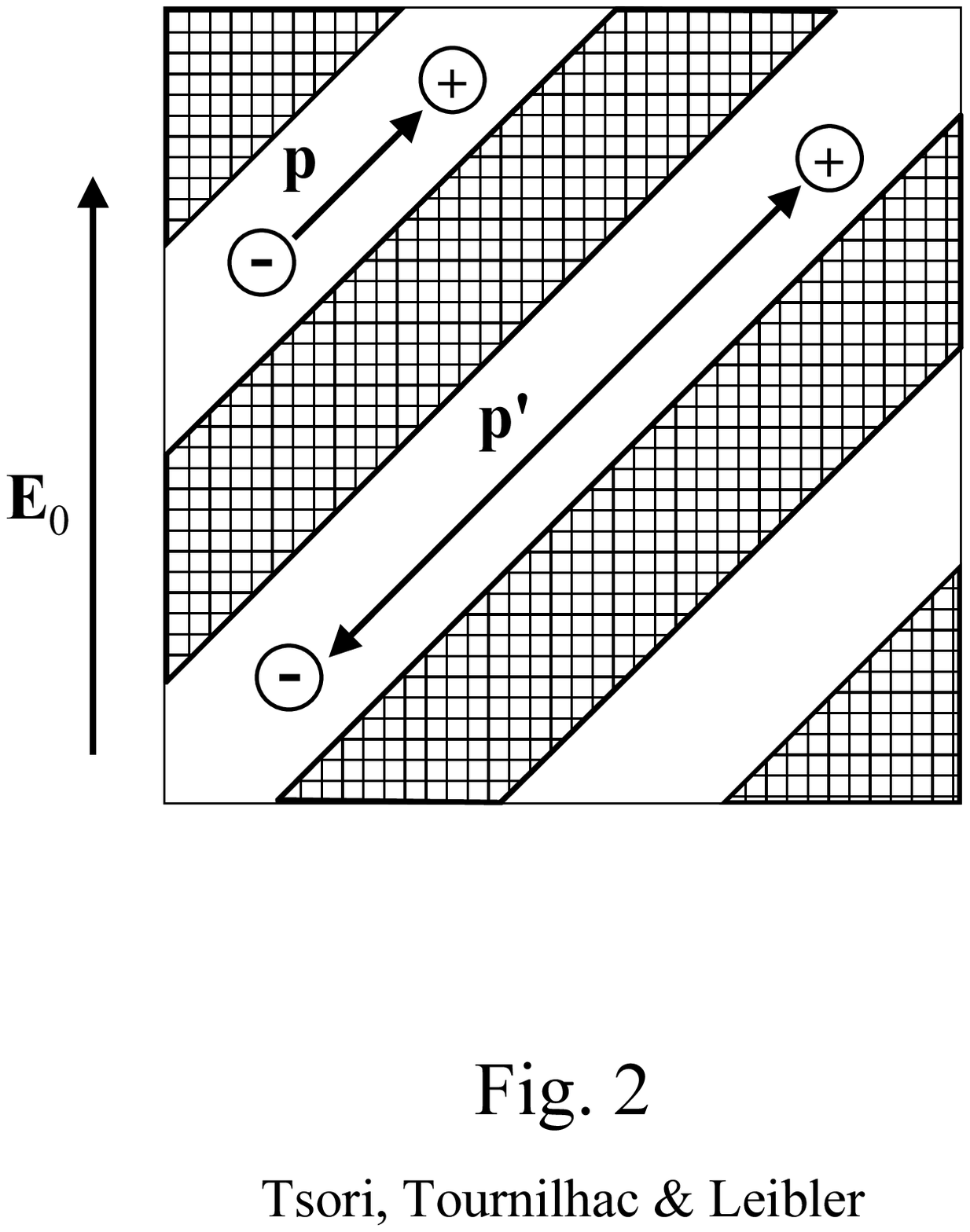}
\end{center}
\caption{Illustration of lamellae in external electric field ${\bf
E}_0$. If the lamellae are tilted with respect to ${\bf E}_0$,
bound-charge dipoles such as ${\bf p}$ cause a torque ${\bf
L}={\bf p}\times{\bf E}_0$. In much the same way, mobile
oscillating ions create an effective dipole ${\bf p'}$ and
contribute a torque ${\bf p}'\times{\bf E}_0$.}
\end{figure}
If the lamellae correspond to alternating stripes of high- and
low-$\eps$, then an induced dipole ${\bf p}$ is created, and a
torque ${\bf p}\times {\bf E}_0$ exists. On the other hand, there
could be materials where $\eps({\bf r})$ is spatially uniform, and
the stripes correspond to high and low conductivity (or mobility)
regions. In an ac field, oscillating ions create an effective
dipole ${\bf p}'$, and a torque ${\bf p}'\times {\bf E}_0$ now
exists.

We calculate first ${\bf L}_{\rm dielec}=\int {\bf p}\times{\bf
E}_0 ~{\rm d}^3r$, where ${\bf p}=\chi({\bf E}_0+{\bf E}_1)$, and
the susceptibility $\chi$ is given by $\eps=1+4\pi\chi$. ${\bf
L}_{\rm dielec}$ can therefore be expressed as ${\bf L}_{\rm
dielec}=\int \chi {\bf E}_1\times{\bf E}_0 ~{\rm d}^3r$. In the
low-frequency limit ${\bf E}_1$ is given by eqs \ref{fourier} and
\ref{lfl_eqn}, and the torque is
\begin{eqnarray}
{\bf L}_{\rm dielec}&=& \sum_{\bf k}\alpha(k,w)\frac{({\bf k}\cdot{\bf
E}_0)}{k^2}{\bf k}\times{\bf E}_0\label{alpha}\\
\alpha(k,w)&=&-\frac{V}{4\pi }\frac{8\pi iw\sigma_0
\mu_k/\mu_0+\eps_kw^2}{8\pi
iw\sigma_0+\eps_0w^2}\eps_k^*~+c.c.\nonumber
\end{eqnarray}
For concreteness, we focus our attention on a simple lamellar
phase in diblock copolymers, consisting of alternating planar
stacks rich in A and B monomers. We see from the above expression
that the torque vanishes if ${\bf k}$ is parallel or perpendicular
to ${\bf E}_0$. This corresponds to lamellae perpendicular or
parallel to the field, respectively. Furthermore, lamellae
parallel to the field are mechanically stable while perpendicular
lamellae are not.

The Helfand and Amundson expression for the torque is recovered in
the absence of free ions, $\rho_0=0$. One has zero conductivity (
$\sigma_0=0$) and
\begin{eqnarray}\label{L_norho}
{\bf L}_{\rm dielec}&=&2\frac{V}{4\pi} \sum_{\bf k}\left[-\frac{
\eps_k\eps_k^*}{\eps_0}\frac{({\bf k}\cdot{\bf E}_0)}{k^2} {\bf
k}\right]\times{\bf E}_0
\end{eqnarray}
Naturally, the mobility $\mu$ does not enter into the expression.

We turn now to calculate the torque due to mobile dissociated
ions. The ionic current has a component along ${\bf E}_0$ and a
component along ${\bf E}_1$. The former does not contribute to the
torque. For each mobile ion, the oscillatory motion along ${\bf
E}_1$ has an average length $\mu eE_1 2/(w\pi)$. The torque is
hence
\begin{eqnarray}
{\bf L}_{\rm ions}&=&\int e^2(\rho^++\rho^-)\mu 2/(w\pi){\bf E}_1
\times{\bf E}_0~{\rm d}^3r
\end{eqnarray}
Using ${\bf E}_1$ from eq \ref{lfl_eqn} and after some algebra we
find that
\begin{eqnarray}
{\bf L}_{\rm ions}&=&\sum_{\bf k}\beta(k,w)\frac{({\bf k}\cdot{\bf
E}_0)}{k^2}{\bf k}
\times{\bf E}_0\label{beta}\\
\beta(k,w)&=&\frac{8 \sigma_0V}{w\pi} \frac{-(8\pi w
\sigma_0/\mu_0)^2\mu_k\mu_k^* -(8\pi w
i\sigma_0/\mu_0)(\mu_k\eps_k^*-\mu_k^*\eps_k)w^2-\eps_k\eps_k^*w^4}
{(\eps_0w^2)^2+64\pi^2w^2\sigma_0^2}\nonumber
\end{eqnarray}
Similar to eq \ref{alpha} for the torque due to fixed
dipoles, $L_{\rm ions}$ vanishes if the lamellae are
parallel or perpendicular to ${\bf E}_0$ (${\bf k}$
perpendicular or parallel to ${\bf E}_0$, respectively).
In the $\sigma_0/w\gg \eps_0$ regime one obtains
\begin{eqnarray}
{\bf L}_{\rm ions}&=&\frac{8 \sigma_0V}{w\pi}\sum_{\bf k}-
\frac{\mu_k\mu_k^*}{\mu_0^2}\frac{({\bf k}\cdot{\bf
E}_0)}{k^2}{\bf k}\times{\bf E}_0\label{L_w20}
\end{eqnarray}
This result is analogous to the Helfand and Amundson
torque, eq \ref{L_norho}. It is shown again that a
dielectric contrast is not necessary in order to orient
lamellae; a {\it mobility} contrast might be enough.
Moreover, from the $1/w$ factor in the above expression,
it is clear that as the field frequency is reduced the
importance of conductive behavior becomes the dominant one.
The expression for the total torque due to fixed and
mobile charges is (eqs \ref{alpha} and \ref{beta})
\begin{eqnarray}\label{L_total}
{\bf L}&=&{\bf L}_{\rm dielec}+{\bf L}_{\rm ions}=\sum_{\bf
k}\left[\alpha(k,w)+\beta(k,w)\right]\frac{({\bf
k}\cdot{\bf E}_0)}{k^2}{\bf k}\times{\bf E}_0
\end{eqnarray}

Let us compare the total torque at a frequency of $50$ Hz (eq
{\ref{L_total}) to the torque due to the dielectric mechanism
only, eq \ref{L_norho}. The ratio between the torques is $L/L_{\rm
dielec}\simeq 1+16(\eps_0^2/\eps_k\eps_k^*)(\mu_k\mu_k^*/\mu_0^2)
(\sigma_0/w\eps_0)$. In the special case where
$\eps_k/\eps_0\simeq \mu_k/\mu_0$ holds, we find that:
\begin{eqnarray}\label{L_ratio}
\frac{L}{L_{\rm dielec}}\simeq
1+16\frac{\sigma_0}{w\eps_0}\simeq 1.4
\end{eqnarray}
This figure emphasizes the importance of free mobile ions
in orientation experiments in BCPs. The effect of ions
vanishes as $w\to \infty$ ($L_{\rm ions}=0$), but for
frequencies smaller than $50$ Hz the mobile ion mechanism
becomes increasingly important and it may even dominate the
orientation process.

\section{Summary}

We consider in this paper orientation of block copolymers in
external electric fields. The results presented here are a
generalization of previous expressions $^{\cite{AH93,AH94,TTAL}}$
for time-varying fields and for ion-containing samples. This is
the realistic situation in many experiments because unless special
measures are taken to extract ions, they will be found in
abundance (e.g., in anionically prepared block copolymers). We
show that as compared to ``clean'' block copolymers, the existence
of dissociated mobile ions can greatly enhance orientation
effects. The underlying physics in these systems is qualitatively
different in several ways. First, depending on the frequency of
applied field and on the mobility, ions undergo oscillatory
movement with amplitude which is larger or smaller than the
feature size of the underlying block copolymer crystal. This
motion is associated with a force which tends to align
microstructures in a specific direction. In the static case the
ionic drift is related to an electrostatic potential energy gain
favoring certain sample morphology.$^{\cite{TTAL}}$

A second change from the no-ions case is that the system has
currents and is dissipative. An estimate for heating shows that
the sample heating should pose no real experimental concern; it
should be relatively simple to control the temperature throughout
the whole sample. We look at forces that generate torque on the
material. It is shown that in the presence of ions these forces
depend on the underlying BCP periodicity. Thus, lamellar block
copolymers of certain chain lengths have different torque from the
same BCP but with twice the chain lengths, in marked contrast to
previous predictions.

In the high frequency domain, ions are almost immobile and
contribute little to the aligning torque. On the other hand, in
the low frequency domain ions exert on the system as a whole
strong aligning torque. These forces can be comparable or even
larger than those due to the ``dielectric mechanism'', as is
exemplified by the ratio of the fields, eq \ref{L_ratio}. Charge
accumulation at the electrodes may be a concern in static fields
(or for very low frequencies), thereby reducing the field in the
sample.$^{\cite{volker1}}$ This problem can be circumvented by
using high enough frequencies where the drift path of the ions
$l_{\rm drift}$ is smaller than the system size $L$, or by using
electrodes permeable to the ions.

It would be interesting to verify the predictions of this paper by
performing an experiment in diblock copolymers with small
dielectric contrast between the two blocks. In the lamellar or
hexagonal mesophases, ac voltage could orient the sample even for
surprisingly small fields. If the BCP is in a bcc phase of spheres
we expect a phase transition from the bcc to the hexagonal phase,
as predicted by us for static dc fields.$^{\cite{TTAL}}$
We hope that this
paper will stimulate further experimental study of block copolymer
orientation in electric fields.

{\bf Acknowledgment} We would like to thank V. Abetz, A. Ajdari,
D. Andelman, M. Cloitre, P. G. de Gennes, A. C. Maggs, T. P.
Russell and T. Thurn-Albrecht for useful discussions. Y.T.
acknowledges support from the Chateaubriand fellowship program.

\end{document}